\documentclass[aps,12pt]{revtex4} 
\usepackage{amsfonts}\usepackage{amsbsy}
\usepackage{mathrsfs}
\usepackage{graphicx}
\def\cal{\mathcal}
\def\lsim{\mathrel{\rlap{
\lower4pt\hbox{\hskip-3pt$\sim$}}
    \raise1pt\hbox{$<$}}}     
\def\gsim{\mathrel{\rlap{
\lower4pt\hbox{\hskip-3pt$\sim$}}
    \raise1pt\hbox{$>$}}}     
\def\vec#1{\mbox{\boldmath $#1$}}
\def\scr#1{\mbox{\scriptsize #1}}
\begin{document}
\title{Equation of state of deconfined matter within dynamical
  quasiparticle description} 
\author{Yu.B.~Ivanov$^{1,2,}$\footnote{e-mail: Y.Ivanov@gsi.de},
V.V. Skokov$^{1,3,}$\footnote{e-mail: V.Skokov@gsi.de}, and 
V.D.~Toneev$^{1,3,}$\footnote{e-mail: V.Toneev@gsi.de}
}

\affiliation{$^1$ Gesellschaft
 f\"ur Schwerionenforschung, Planckstr.$\!$ 1,
64291 Darmstadt, Germany
\\ $^2$ Kurchatov Institute, Kurchatov
sq.$\!$ 1, Moscow 123182, Russia
\\
$^3$ Joint Institute for Nuclear Research,
 141980 Dubna, Moscow Region, Russia} 

\begin{abstract}
A simple quasiparticle model, motivated by lowest-order perturbative
QCD, is proposed. It is applied to interpret the lattice QCD  equation
of state. A reasonable reproduction of the lattice data is
obtained. In contrast to existing quasiparticle models, the present
model is formulated in dynamical rather than thermodynamical terms,
and is easily applicable to a system with finite baryon density. 
In particular, the model simulates the
confinement property. \\[2mm] 
\end{abstract}
\maketitle


\section{Introduction}

The most fundamental way to compute properties of strongly interacting
matter and, in particular, its equation of state (EoS) is provided by
lattice QCD calculations \cite{latEoS0}. The technique of these
calculations rapidly progresses. Recently, lattice data on the EoS at
finite baryon chemical potential became available
\cite{latEoS,Allton,Gavai}. 
Interpretation of these data within the {\em straightforward} QCD
perturbation theory \cite{Arnold} 
is hardly possible in view of its extremely poor convergence for any
temperature of practical interest. To overcome this poor convergence,
resummation schemes were developed. A scheme 
based on hard-thermal-loop (HTL) effective action
\cite{Pisarski} 
have been proposed, with alternative formulations in the form of
so-called HTL perturbation theory \cite{Andersen} or based on the
$\Phi$-derivable approximation \cite{Blaizot}. This approach
justified a picture of weakly interacting quasiparticles, as
determined by the HTL propagators, and resulted in remarkably good
agreement with lattice data above $3T_C$, with $T_C$ being the
critical temperature of the phase transition. It is important to
emphasize that this quasiparicle picture emerges directly from the QCD
dynamics, although treated within the thermal framework. 
Recently, a new resummation scheme based on dimensionally reduced
screened perturbation theory (DRSPT) was proposed
\cite{Blaizot03}. In certain sense, the efficiency of this DRSPT
scheme even surpasses that of the HTL perturbation theory. However, it
still gives reliable results only above $2.5T_C$.

To extend this perturbative description below $3T_C$, various
phenomenological quasiparticle models 
\cite{Gorenstein95,Greiner,Levai,Pesh96,Szabo03,Rebhan03,Weise01} 
were proposed. These models are formulated in terms of massive quarks
and gluons and are constructed in such a way that at high
temperatures they match the perturbative results and then extend them
down in temperature. 
It is not clear, if a quasiparicle picture is relevant
below $3T_C$ at all. Therefore, all these models are purely
phenomenological. Nevertheless, interpretation of lattice data within
these quasiparticle models turned out to be very
successful. With few phenomenological parameters it was possible to
reasonably reproduce all lattice thermodynamic quantities. The feature
of all above cited quasiparticle models, which still looks slightly
irritating, is that they are formulated in terms of thermodynamic
quantities (i.e. temperature $T$ and baryon chemical potential $\mu$)
rather than dynamical ones, like various densities. From the
theoretical point of view, the quasiparticle picture should be
formulated in dynamical terms. The thermal equilibrium is only a
particular case of this general picture. From the practical point of
view, if the quasiparticle is introduced as a dynamical object, it
would be possible  
to use a quasiparticle model for extending the equilibrium lattice
description to (at least, slightly) nonequilibrium configurations,
relying on reasonable reproduction of equilibrium properties by this
model. Such kind of extension is really required for analyzing
heavy-ion collisions, where the thermalization is still a debating
problem.  

In the present paper we propose a simple quasiparticle model
formulated in dynamical terms. In construction of this model, we proceed
from properties of the perturbative solution to QCD, justifying  the
quasiparticle picture, rather than from first principles of QCD. 

\section{Dynamical Quasiparticles}

Here we will follow the line of
Refs. \cite{Gorenstein95,Pesh96,Szabo03,Rebhan03}, assuming only
massive quasiparticles and avoiding artificial reduction of
quark--gluon degrees of freedom as in \cite{Weise01}.  
Let the effective Lagrangian for transverse gluons $\phi_a$ and
quarks $\psi_{qc}$ of $N_f$ flavors be as follows 
\begin{eqnarray}
\label{qg-lagr}
{\cal L} &=& \frac{1}{2}\sum_{a=1}^{N_g}
\left((\partial_\mu \phi_a)^2 - m_g^2(\eta,{\vec\xi}) \phi_a^2\right)
\cr &+&
\sum_{c=1}^{N_c}\sum_{q=1}^{N_f}
\bar{\psi}_{qc} [i\gamma_\mu \partial^\mu 
- m_q(\eta,{\vec\xi})]\psi_{qc}
-B(\eta,{\vec\xi}), 
\end{eqnarray}
where $N_c=3$ is number of colors, $N_g=2(N_c^2-1)$ is number of
transverse gluons, taking into account two transverse polarizations,   
$m_g$ and $m_q$
are effective masses of gluons and quarks, respectively, depending on 
self-consistent fields  $\eta$ and ${\vec\xi}= \{\xi_1,...,\xi_{N_f}\}$. 
$B(\eta,{\vec\xi})$ is a potential of mean-field self-interaction. 
Writing down Lagrangian (\ref{qg-lagr}) we have omitted kinetic terms
of the $\eta$ and ${\vec\xi}$ fields, assuming that they are not
essential for the problem. Note that these kinetic terms are precisely
zero in the spatially homogeneous equilibrium and hence are really
negligible for slight deviations from it. This Lagrangian  is written
proceeding from general features of the perturbative solution to QCD,
which claims that quarks and transverse gluons are weakly interacting
quasiparticles. Here all interactions between gluons and quarks, as
well as their 
self-interactions, are hidden in their effective  masses depending on
mean fields, which in their turn are determined in terms of these
masses.

Equations of motion for the mean fields are derived
in the standard way:  
\begin{eqnarray}
\label{qg-gap-1}
-\frac{\partial B}{\partial\eta^2}&=& 
\frac{1}{2}\frac{\partial m_g^2}{\partial\eta^2} 
\sum_{a=1}^{N_g}\left\langle \phi_a^2\right\rangle
+
\frac{1}{2}\sum_{q=1}^{N_f}
\frac{\partial m_q^2}{\partial\eta^2}
\sum_{c=1}^{N_c}
\frac{\left\langle \bar{\psi}_{qc}\psi_{qc}\right\rangle}{m_q}
, 
\\
\label{qg-gap-2}
-\frac{\partial B}{\partial\xi_i^2}&=& 
\frac{1}{2}\frac{\partial m_g^2}{\partial\xi_i^2} 
\sum_{a=1}^{N_g}\left\langle \phi_a^2\right\rangle
+
\frac{1}{2}\sum_{q=1}^{N_f}
\frac{\partial m_q^2}{\partial\xi_i^2}
\sum_{c=1}^{N_c}
\frac{\left\langle \bar{\psi}_{qc}\psi_{qc}\right\rangle}{m_q}
. 
\end{eqnarray}
Here 
\begin{eqnarray}
\label{g-rho1}
\sum_{a=1}^{N_g}\left\langle \phi_a^2\right\rangle&=&\frac{N_g}{2\pi^2} 
\int_0^\infty \frac{k^2\;dk}{(k^2+m_g^2)^{1/2}} f_g (x,k), 
\\
\label{q-rho1}
\sum_{c=1}^{N_c}
\frac{\left\langle \bar{\psi}_{qc}\psi_{qc}\right\rangle}{m_q}&=&
\frac{N_c}{\pi^2} 
\int_0^\infty \frac{k^2\;dk}{(k^2+m_q^2)^{1/2}} [f_q (x,k)+f_{\bar{q}} (x,k)]
\end{eqnarray}
are scalar densities of gluons and quarks divided by mass, respectively, 
with $f_g (x,k)$, $f_q (x,k)$ and $f_{\bar{q}} (x,k)$ being distribution
functions of gluons, quarks and antiquarks in
space--time ($x$) and 4-momenta ($k$). In the particular case of
thermal equilibrium we going to consider here, these are  
\begin{eqnarray}
\label{fg}
f_g (k)&=&\frac{1}{\exp[(k^2+m_g^2)^{1/2}/T]-1},
\\
\label{fq}
f_q (k)&=&\frac{1}{\exp\{[(k^2+m_q^2)^{1/2}-\mu_q]/T\}+1},
\\
\label{faq}
f_{\bar{q}} (k)&=&\frac{1}{\exp\{[(k^2+m_q^2)^{1/2}+\mu_q]/T\}+1}.
\end{eqnarray}
where $T$ is the temperature, and $\mu_q$ is the $q$-quark chemical
potential. In general, all $\mu_q$ may be different.  
If we consider a system with zero overall strangeness and charm,
$\mu_q$ relates to the baryon chemical potential as
$\mu_u=\mu_d=\mu/3$ with all other $\mu_q=0$. 

A solution of Eqs (\ref{qg-gap-1}) and (\ref{qg-gap-2}), which are
usually referred as gap equations, provides us with an expressions of 
fields $\eta$ and $\xi_q$ in terms of above scalar densities. 
Without loosing generality,
it is convenient to demand that these solutions for the $\eta$ and
${\vec\xi}$ fields are given by scalar 
densities of gluons and quarks divided by mass
\begin{eqnarray}
\label{eta}
\eta^2&=&\sum_{a=1}^{N_g}\left\langle \phi_a^2\right\rangle,
\\
\label{xi}
\xi_q^2&=&\sum_{c=1}^{N_c}
\frac{\left\langle \bar{\psi}_{qc}\psi_{qc}\right\rangle}{m_q}. 
\end{eqnarray}
Indeed, had we started
  from other collective variables $\widetilde{\eta}$ and
  $\widetilde{\vec\xi}$, which differ from  $\eta$ and
${\vec\xi}$ defined by Eqs (\ref{eta}) and (\ref{xi}), and the
corresponding potential
$\widetilde{B}(\widetilde{\eta},\widetilde{\vec\xi})$, equations
of motion (\ref{qg-gap-1}) and (\ref{qg-gap-2}) would provide us with
solutions $\widetilde{\eta}(\eta,{\vec\xi})$ and 
  $\widetilde{\vec\xi}(\eta,{\vec\xi})$ with $\eta$ and
${\vec\xi}$ associated with densities (\ref{eta}) and (\ref{xi}). Then
we could immediately redefine the potential as 
$B(\eta,{\vec\xi})=\widetilde{B}(\widetilde{\eta}(\eta,{\vec\xi}),\widetilde{\vec\xi}(\eta,{\vec\xi}))$
and thus transform to desired variables $\eta$ and ${\vec\xi}$.


The high-temperature limit, $T \gg T_C$, where $T_C$ is the temperature
of the deconfinement phase transition, puts certain constrains on the
functional dependence of the effective masses $m_g(\eta,{\vec\xi})$
and $m_q(\eta,{\vec\xi})$. In this limit, the straightforward
calculation of above scalar densities in the leading order results in  
\begin{eqnarray}
\label{g-rho-1}
\eta^2 (T \gg T_C) &\simeq& \frac{N_g}{12} T^2, 
\\
\label{q-rho-1}
\xi_q^2 (\mu_q, T \gg T_C) 
&\simeq& N_c\left(\frac{1}{6} T^2 + \frac{1}{2\pi^2} \mu_q^2\right). 
\end{eqnarray}
Perturbative values of $m_g$ and $m_q$  are also known \cite{Lebel96}
\begin{eqnarray}
\label{m_g-qg-T}
m_g^2 (\{\mu_q\},T \gg T_C) &=&  
\frac{1}{12}\left( (2N_c+N_f^{\scr{eff}})T^2 + 
\frac{3}{\pi^2}\sum_{q=1}^{N_f^{\scr{eff}}}
\mu_q^2\right)g^2(\{\mu_q\},T \gg T_C),
\\
\label{m_q-qg-T}
m_q^2(\{\mu_q\},T \gg T_C) -m_{q0}^2 &=&  
\frac{N_g}{16N_c}\left(T^2+\frac{\mu_q^2}{\pi^2}\right)g^2(\{\mu_q\},T \gg T_C),
\end{eqnarray}
where $m_{q0}$ is the current mass of the $q$-quark, $N_f^{\scr{eff}}$
is the effective number of quark flavors which can be excited, and
$g^2$ is the QCD running coupling constant squared, generally
depending on $T$ and all $\mu_q$. 
In the particular case of all
$\mu_q=0$, the latter is also known \cite{Yndurain,Kapusta}, in the 2-loop
approximation it is 
\begin{eqnarray}
\label{g-0}
g^2(\{\mu_q=0\},T \gg T_C)=\frac{16\pi^2}{ \beta_0\ln(2\pi T/\Lambda)^2}
\left(
1 - \frac{2\beta_1}{\beta_0^2}
\frac{\ln\ln(2\pi T/\Lambda)^2}{\ln(2\pi T/\Lambda)^2}
\right)
\end{eqnarray}
with 
\begin{eqnarray}
\label{beta}
\beta_0 = \frac{1}{3}(11 N_c - 2 N_f^{\scr{eff}}), \quad
\beta_1 = \frac{1}{6}(34 N_c^2 - 13 N_c N_f^{\scr{eff}} 
+ 3 N_f^{\scr{eff}}/N_c), 
\end{eqnarray}
and $\Lambda$  being the QCD scale.  
The energy scale is taken here 
equal to $2\pi T$, i.e. the first nonzero Mastubara frequency. In HTL
calculations this scale is sometimes varied from $\pi T$ to $4\pi T$
to determine theoretical error bars.

Expressing $T$ and all $\mu_q$ in terms of scalar densities $\eta$ and
${\vec\xi}$ of Eqs. (\ref{g-rho-1}) and (\ref{q-rho-1}), and
substituting them into expressions for asymptotic effective masses
(\ref{m_g-qg-T}) and (\ref{m_q-qg-T}), we arrive at the following
expressions for the latter  
\begin{eqnarray}
\label{m_g-qg}
m_g^2 (\eta,{\vec\xi}) &=&  
\left(\frac{2N_c}{N_g} \eta^2 +  \frac{1}{2N_c} \sum_{q=1}^{N_f}\xi_q^2
\right)g^2(\eta^2,{\vec\xi}), 
\\
\label{m_q-qg}
m_q^2 (\eta,{\vec\xi})-m_{q0}^2 &=&  
\left(\frac{1}{2N_c} \eta^2 + \frac{N_g}{8N_c^2} \xi_q^2
\right)g^2(\eta^2,{\vec\xi}). 
\end{eqnarray}
Here we keep $N_f$ instead of $N_f^{\scr{eff}}$, since quark densities
$\xi_q^2$ automatically take care of quark contribution,
i.e. $\xi_q^2\approx 0$, if the $q$-quark is too heavy. 
These expressions give at the same time the ansatz for
the effective masses in terms of scalar densities.

Now our goal is to find an appropriate expression for
$g^2(\eta,{\vec\xi})$, which takes the limit (\ref{g-0}) at
$\mu_q=0$ and $T\gg T_C$, and then to solve gap equations
(\ref{qg-gap-1}) and (\ref{qg-gap-2}) with respect to
$B(\eta,{\vec\xi})$. Then the quasiparticle model would be completely
defined, and we could do any calculations both in thermodynamics and
nonequilibrium. Unfortunately, we failed to solve this problem in
general, i.e. we have not found an appropriate potential $B$ which is
required for calculation of the thermodynamic quantities, cf. Eqs
(\ref{E_qg-s}) and (\ref{P_qg-s}). 
However, we have found an elegant solution in the particular case,
when number of colors equals number of flavors, i.e. $N_f=N_c$. In fact,
this case is quite general for comparison to lattice data as well as
for possible applications in astrophysics and heavy-ion physics. 

\section{Particular Case of $N_f=N_c$}

Let us consider a particular case when $N_f=N_c$ while the quarks may
be different, i.e. their current masses $m_{q0}$ as well as chemical
potentials $\mu_q$ may differ.

Since $N_f=N_c$, Eqs. (\ref{m_g-qg}) and (\ref{m_q-qg}) can be represented as
follows 
\begin{eqnarray}
\label{m_g-qg-s}
m_g^2  &=&  
\frac{2}{N_g} \sum_{q=1}^{N_f} \zeta_q^2 g^2(\chi),
\\
m_q^2-m_{q0}^2 &=&  
\frac{1}{2N_c} \zeta_q^2 g^2(\chi),
\label{m_q-qg-s}
\end{eqnarray}
where
\begin{eqnarray}
\label{zeta}
\zeta_q^2&=&\eta^2 + \frac{N_g}{4N_c}\xi_q^2, 
\\
\label{chi}
\chi^2 &=& \left(\sum_{q=1}^{N_f} \zeta_q^4\right)^{1/2}.  
\end{eqnarray}
Here we just guessed that the $\chi$ dependence of $g^2$ is the proper
one. This functional dependence is required to define the potential
$B$, with which gap equations (\ref{qg-gap-1}) and (\ref{qg-gap-2})
give solutions for masses precisely in the form of Eqs (\ref{m_g-qg-s}) and
(\ref{m_q-qg-s}). Indeed, gap Eqs. (\ref{qg-gap-1}) and (\ref{qg-gap-2})
in terms of new variables $\zeta_q^2$ read
\begin{eqnarray}
\label{qg-gap-s}
-\frac{\partial B}{\partial\zeta_i^2}&=& 
\frac{1}{2}\frac{\partial m_g^2}{\partial\zeta_i^2} \eta^2 
+
\frac{1}{2}\sum_{q=1}^{N_f}
\frac{\partial (m_q^2-m_{q0}^2)}{\partial\zeta_i^2}
\xi_q^2
\cr
&=&
\frac{1}{N_g}\left(g^2 + 
\frac{d g^2}{d \chi^2}\frac{\zeta_i^2}{\chi^2}\sum_{q=1}^{N_f}\zeta_q^2
\right)\eta^2
+
\frac{1}{4N_c}\sum_{q=1}^{N_f}\left(
\delta_{qi} g^2 +
\zeta_q^2\frac{d g^2}{d \chi^2}\frac{\zeta_i^2}{\chi^2}
\right)\xi_q^2
\cr
&=&
\frac{1}{N_g}g^2\left(
\eta^2 + \frac{N_g}{4N_c}\xi_i^2\right)
+
\frac{1}{N_g}
\frac{d g^2}{d \chi^2}\frac{\zeta_i^2}{\chi^2}
\sum_{q=1}^{N_f}\zeta_q^2\left(\eta^2 + \frac{N_g}{4N_c}\xi_q^2\right)
\cr
&=&
\frac{1}{N_g}\left(
g^2 + \chi^2\frac{d g^2}{d \chi^2}\right)\zeta_i^2. 
\end{eqnarray}
where $i=u,d$ or $s$. 
In fact, this is the main trick advanced by the peculiar case of $N_f=N_c$.
This relation implies that 
the potential of mean-field self-interaction
$B(\eta,{\vec\xi})$ is in fact a function of a single variable
$\chi$, and gap equations (\ref{qg-gap-1}) and (\ref{qg-gap-2}) are
reduced to the single one 
\begin{eqnarray}
\label{qg-gap-s2}
\frac{d B}{d \chi}=
-\frac{1}{N_g} \chi^2\frac{d (\chi^2 g^2)}{d \chi}, 
\end{eqnarray}
integration of which is straightforward 
\begin{eqnarray}
\label{U_g}
B(\chi)
&=& B_C-
\frac{1}{N_g}\left[\chi^4 g^2(\chi)-\chi_C^4 g^2(\chi_C)\right]
+
\frac{2}{N_g}\int_{\chi_C}^{\chi} d\chi_1 \chi_1^3g^2(\chi_1)
\end{eqnarray}
with $B_C$ being an integration constant. In fact, this
function $B(\chi)$ has a meaning of the bag constant of the bag
model. However, we will refer $B_C$ as the ``bag parameter'',
since it is really constant. 

Thus, we succeeded to determine the
$B$ potential, which is required for thermodynamically consistent
calculation of thermodynamic quantities
(\ref{E_qg-s})--(\ref{n_qg-s}). We cannot claim that this is the only
possible solution for this $B$, because it was obtained as a result of
certain guess. However, this solution works quite well in reproducing
lattice data, as it is demonstrated in the next section.

A reasonable ansatz for the coupling constant itself is as follows 
\begin{eqnarray}
\label{g2_eff}
g^2(\chi)  =  
\frac{16\pi^2}{ \beta_0\ln[(\chi^2+\chi_0^2)/\chi_C^2]}
f(\chi), 
%
\end{eqnarray}
where $\chi_C^2$ and $\chi_0^2$ are some phenomenological parameters,
and an auxiliary function $f(\chi)$, meeting the condition
$f(\chi\to\infty)\to 1$, helps us to choose between 1-loop
($f(\chi)\equiv 1$) and 2-loop asymptotics of the coupling constant,
cf. Eq. (\ref{g-0}). Two reasonable choices of this auxiliary function
are discussed below, in sect. \ref{Comparison}.  
This $g^2(\chi^2)$ indeed takes the limit (\ref{g-0}) 
at $\mu_q=0$ and $T\gg T_C$, provided $\chi_0^2\ll \chi_C^2$ and {\em
  properly} defined $\chi_C^2$ in terms of $\Lambda$. The {\em proper}
definition in the case, when the temperature is much larger than all
current quark masses, $T\gg m_{q0}$, is as follows   
$$\chi_C^2=\frac{N_g N_c^{1/2}}{8}  
\left(\frac{\Lambda}{2\pi}\right)^2.$$

It is appropriate to mention here that in spite of the declared case
$N_f=N_c$, we are able to consider less number of flavours,
$N_f^{\scr{eff}}<3$, cf. Eq. (\ref{beta}), within the same
formalism. To exclude a $q$-quark flavor from the treatment at certain
temperature $T$, we should take its current mass to be large: 
$m_{q0} \gg T$, which implies $m_{q} \gg T$,
cf. Eq. (\ref{m_q-qg-s}). In this limit the respective density
$\xi_q^2 \to 0$ 
and simply falls out of the corresponding $\zeta_q^2$,
cf. Eq. (\ref{zeta}), and hence out of the calculation scheme for 
lighter particles. Taking into account that the contribution of this
heavy quark into thermodynamic quantities 
(\ref{E_qg-s})--(\ref{n_qg-s}) is negligible as compared with
that of lighter particles, we see that this heavy quark turns out to
be completely switched off from the calculation, as if it does not
exist. If we consider $m_{q0} \gg T \gg T_C$, the contribution of this
heavy quark disappears even from asymptotic formulas (\ref{m_g-qg-T})
and (\ref{m_q-qg-T}). However, the delicate feature of the present
solution is that we still should keep the flavor summation in  
Eq. (\ref{m_g-qg-s}) running though all 3 flavors, in order to obtain
the proper gluon contribution into the gluon mass, even in the case of 
$m_{q0} \gg T \gg T_C$. The reason is that for the heavy quark we still
have $\zeta_q^2=\eta^2$, i.e. the gluon density which is
nonzero. Thus, e.g. for the 
2-flavour case, we should keep $N_f=3$ whereas take $N_f^{\scr{eff}}=2$
in Eq. (\ref{beta}) required for definition of the coupling constant
(\ref{g2_eff}).


To summarize, the procedure of solving the model equations is as
follows. First, we define all the free parameters of the model
($\chi_C$, $\chi_0$, $B_C$), including the auxiliary function
$f(\chi)$. Given the temperature $T$ and the set of chemical
potentials $\mu_q$, implicit set of equations (\ref{g-rho1}),
(\ref{q-rho1}), (\ref{m_g-qg-s})--(\ref{chi}) and (\ref{g2_eff})
should be solved. 
As a result of this solution, we obtain effective quark and gluon
masses and the value of $\chi$ variable, which is required for
calculation of $B(\chi)$, cf. Eq. \ref{U_g}).  
Now, when all the quantities are defined, we can calculate the energy
density $\varepsilon(T,\mu)$, pressure $P(T,\mu)$, and baryon density
$n_B(T,\mu)$ as follows  
\begin{eqnarray}
\label{E_qg-s}
\hspace*{-9mm}
\varepsilon(T,\mu)&=&\frac{N_g}{2\pi^2} 
\int_0^\infty k^2\;dk\;(k^2+m_g^2)^{1/2}\; f_g (k) 
\cr &&
+\sum_{q=1}^{N_f}
\frac{N_c}{\pi^2} 
\int_0^\infty k^2\;dk\;(k^2+m_q^2)^{1/2}\; [f_q (k)+f_{\bar{q}} (k)]
+B(\chi),
\\
\label{P_qg-s}
\hspace*{-9mm}
P(T,\mu)&=&\frac{N_g}{6\pi^2} 
\int_0^\infty \frac{k^4\;dk}{(k^2+m_g^2)^{1/2}} f_g (k)
\cr &&
+\sum_{q=1}^{N_f}
\frac{N_c}{3\pi^2} 
\int_0^\infty \frac{k^4\;dk}{(k^2+m_q^2)^{1/2}} [f_q (k)+f_{\bar{q}} (k)]
-B(\chi), 
\\
\label{n_qg-s}
\hspace*{-9mm}
n_B(T,\mu)&=&\frac{1}{3}\sum_{q=1}^{N_f}
\frac{N_c}{\pi^2} 
\int_0^\infty k^2\;dk\; [f_q (k)-f_{\bar{q}} (k)].
\end{eqnarray}
Note that the thermodynamic consistency is automatically fulfilled in
this scheme, since we proceed from a proper Lagrangian formulation.

In particular, we would like to mention that the present model
simulates the confinement of quarks and gluons. When temperature
and/or chemical potentials decrease, the densities, $\eta^2$ and
$\xi_q^2$, and together with them the variable $\chi$ drop down.
At some value of $\chi$ the argument of
$\ln[(\chi^2+\chi_0^2)/\chi_C^2]$ in Eq. (\ref{g2_eff}) becomes very
close to 1, and hence $g^2\to\infty$. Thus, there are no solutions to
the above equations below certain values of temperature and chemical
potentials. This can be interpreted as a kind of confinement.

\section{Comparison with Lattice Data} 
\label{Comparison}

Our goal is to fit the above described model to the recent (2+1)
flavour lattice data for nonzero chemical potentials
\cite{latEoS}. To be consistent with these lattice data, we accepted
current quark masses $m_{u0}=m_{d0}=$ 65 MeV and $m_{s0}=$ 135 MeV,
which were used in these lattice calculations. As we have found out,
the actual results of our quasiparticle model are quite insensitive to
variation of $m_{q0}$ from above lattice values to the ``physical''
ones $m_{u0}=m_{d0}=$ 7 MeV and $m_{s0}=$ 150 MeV. The model also
involves several phenomenological parameters: the ``bag parameter''
$B_C$, cf. (\ref{U_g}), the ``QCD scale'' $\chi_C$ and  an auxiliary
function $f(\chi)$, cf. (\ref{g2_eff}). 
Another parameter $\chi_0^2$,
as it was expected, should be taken small $\chi_0^2\ll\chi_C^2$.
In fact, it shifts the lower limit of integration in the
expression for $B(\chi)$, cf. Eq. (\ref{U_g}), from the singular point
of the coupling constant,  cf. Eq. (\ref{g2_eff}), and hence
regularizes the calculation of $B(\chi)$. Therefore, it is closely
related to the ``bag parameter'' $B_C$, which is an integration
constant in the same expression. A change of $\chi_0^2$
implies the corresponding change of $B_C$. In
all further calculations we take $\chi_0^2= 0.01 \chi_C^2$, and hence
the below stated values of $B_C$ correspond only to this choice.

An implicit parameter
of our model is the critical temperature $T_C$, i.e. the temperature
at which the deconfinement phase transition occurs at $\mu=$ 0. We
could identify this temperature with that of the end point of the 
solution discussed above. However, this end point is numerically
determined not quite reliably because of the singular behavior of the
solution near it. Another reason is that the end-point temperature
should not necessary coincide with $T_C$. The phase transition at
$\mu=$ 0 in the case of (2+1) flavours is of the cross-over type, as
it was found in lattice calculations. This implies that a strong
interplay between quark--gluon and hadronic degrees of freedom occurs
near $T_C$, which actually determines the $T_C$ value itself. As we
completely disregard the hadronic degrees of freedom in the model, we
cannot count on proper determinations of $T_C$ value. Therefore, we
vary $T_C$ from the determined end-point temperature to slightly below
in order to achieve the best fit of the lattice data.  

\begin{figure}[ht]
\begin{center}
\includegraphics[height=64mm,clip]{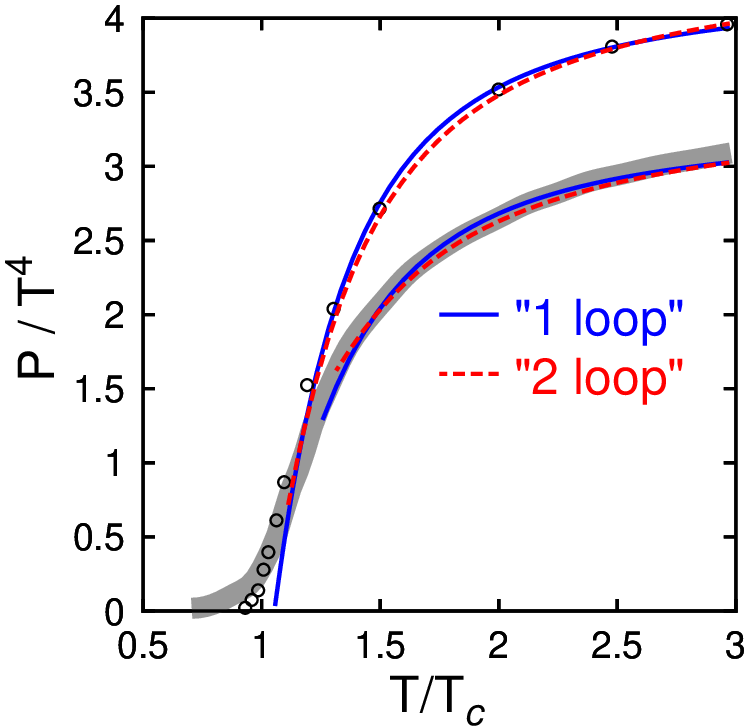}
\end{center}
\caption{Pressure normalized by $T^4$ as a function of $T/T_c$ at
  $\mu=0$. The solid line corresponds to ``1-loop'' calculation with
  the overall normalization factor of 0.9. The dashed line represents
  the ``2-loop'' calculation. The (2+1)-flavour lattice data
  \cite{latEoS} are displayed by open circles, and the 2-flavour
  lattice data \cite{latEoS0} -- by grey band.} 
\label{p0_fig}
\end{figure}

As for the auxiliary function $f(\chi)$, our first choice was 
\begin{eqnarray}
\label{1-loop}
f_{\scr{1-loop}}(\chi)\equiv 1, 
\end{eqnarray}
which we refer as ``1-loop'' choice, because with this
$f_{\scr{1-loop}}$ the coupling constant takes the 1-loop perturbative
limit at $T\to\infty$, cf. Eq. (\ref{g-0}). In this case we are left
with only two basic parameters,  $\chi_C$ and $B_C$. These are fitted
to reproduce the form of the pressure as a function of temperature at
zero chemical potential. 
However, these two parameters does not allow us to reproduce the
overall normalization of the lattice pressure. With this respect, it
is suitable to recollect that the overall normalization of the lattice
data is somewhat uncertain. Indeed, the lattice calculations were done
on lattices with $N_t=$ 4 temporal extension \cite{latEoS}. To
transform the raw lattice data into physical ones, i.e. to extrapolate
to the continuum case of $N_t\to\infty$, the raw data are multiplied
by {\em ``the dominant $T\to\infty$ correction factors between the
  $N_t=$ 4 and continuum case''}, $c_p=$ 0.518 and $c_\mu=$ 0.446,
\cite{latEoS}. These factors are determined as a ratios of the
Stefan--Boltzmann pressure at $\mu=$ 0 ($c_p$) and the $\mu$-dependent
part of the Stefan--Boltzmann pressure ($c_\mu$) to the corresponding
values on the $N_t=$ 4 lattice \cite{latEoS}. In view of this
uncertainty, it is legal to apply {\em an additional overall
  normalization factor} to the same quantities calculated within
quasiparticle model \footnote{It would be more reasonable to apply this
  additional normalization factor to the lattice data. However, we do
  not want to distort the ``experimental'' results.}. In order to keep
the number of fitting parameters as few as possible, we use a single
normalization factor instead of two different ones, $c_p$  and
$c_\mu$, in the lattice calculations. For the best fit of
the lattice data the overall normalization factor was chosen equal 0.9 and
$T_C$ was shifted slightly below the end-point temperature, which by
itself was determined quite approximately.  
Note that the fitted $T_C=$ 195 MeV is slightly above its lattice
value 175 MeV.  
The set of parameters is summarized in Table. The result of the fit is
presented in Fig. \ref{p0_fig}. In the same figure, also comparison
with 2-flavour lattice data \cite{latEoS0} is presented. For the present
``1-loop'' variant, 2-flavour data are perfectly reproduced with the
same set of parameters as for (2+1)-flavour case, only the current
mass of the strange quark was taken $m_{s0}=$ 100 GeV in order to suppress its
contribution. In this case the result for critical temperature is even
better: $T_C^{\scr{(2f)}}=$ 175 MeV, which well complies with its lattice
value \cite{latEoS0}.

\begin{center}
\hspace*{-9mm}
\begin{minipage}[t]{140truemm}
\begin{center}
\vspace*{2mm}
  \begin{tabular}{|c|c|c|c|c|c|c|c|}
\hline
 Version& $N_f^{\scr{eff}}$$^*$ & $T_C$, MeV & $\chi_C$, MeV &  
$B_C/\chi_C^4$$^{**}$ 
& $f$-factor$^{***}$  & overall normalization factor\\ \hline 
\hline
 ``1-loop''& 2+1          & 195 & 141.3 & -97.5  & 1 & 0.9  \\ \hline
 ``1-loop''& 2$^{****}$   & 175 & 141.3 & -97.5  & 1 & 0.9  \\ \hline
 ``2-loop''& 2+1          & 195 & 119.6 & -267.5 & 2.6 & 1  \\ \hline
 ``2-loop''& 2$^{****}$   & 175 & 119.6 & -262.0 & 2.6 & 1  \\ \hline
   \end{tabular}
\\
\end{center}
$^*$ {\small Please, do not confuse it with $N_f$, which should be
  always $N_f=3$.}\\
$^{**}$ {\small These $B_C$ values correspond to the $\chi_0^2=
  0.01 \chi_C^2$ choice.}\\
$^{***}$ {\small This is the effective value of the auxiliary function
  $f(\chi)$ in the temperature range under investigation, i.e. from
  $T_C$ to $3T_C$.}\\   
$^{****}$ {\small $m_{s0}=$ 100 GeV in order to suppress the $s$-quark
contribution.}   
\\[-5mm]
\begin{center}
{\bf Table:} Best fits of quasiparticle parameters to the lattice data
\cite{latEoS0,latEoS}. 
\vspace*{3mm}
\end{center}
\end{minipage}
\end{center}

\begin{figure}[h]
\begin{center}
\begin{minipage}[h]{65mm} 
\includegraphics[height=64mm,clip]{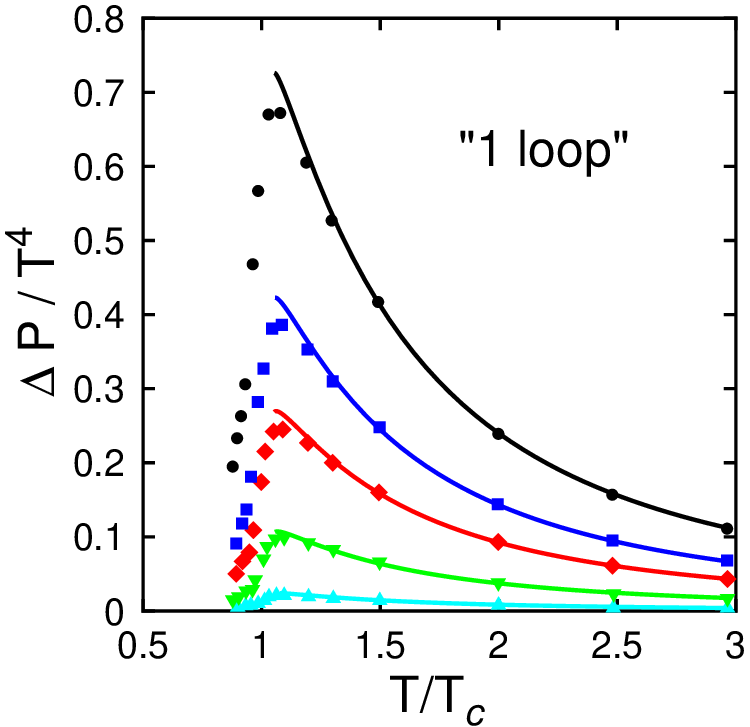}
\end{minipage}
\hspace*{9mm}
\begin{minipage}[h]{65mm} 
\includegraphics[height=64mm,clip]{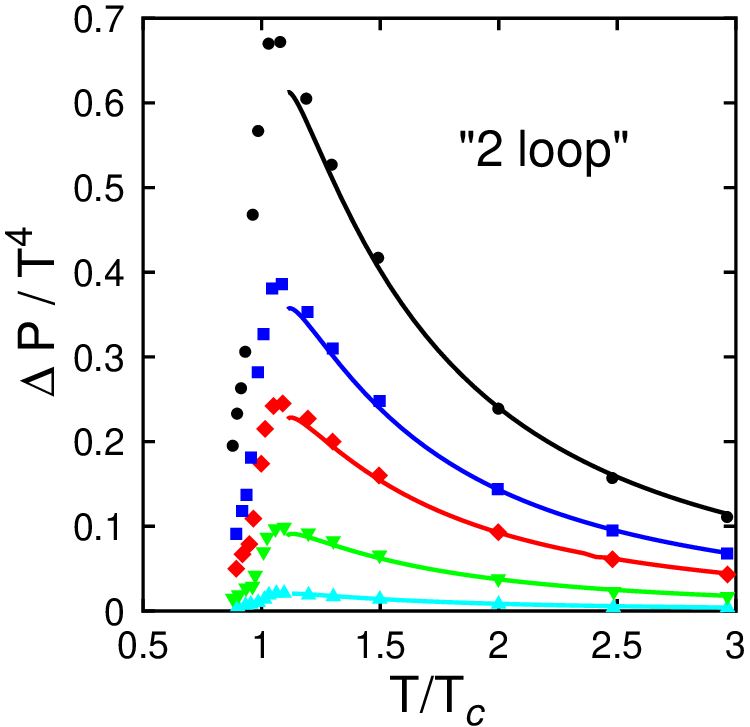}
\end{minipage}
\end{center}
\caption{$\Delta P = P(T,\mu)-P(T,\mu=0)$ normalized by $T^4$ as a
  function of $T/T_c$ at $\mu =$ 100, 210, 330, 410, 530 MeV (from
  bottom to top) within the  ``1-loop'' (left panel) and
  ``2-loop'' (right panel) calculations. The (2+1)-flavour lattice
  data are from \cite{latEoS}. 
} \label{dp}
\end{figure}
\begin{figure}[h]
\begin{center}
\begin{minipage}[h]{65mm} 
\includegraphics[height=64mm,clip]{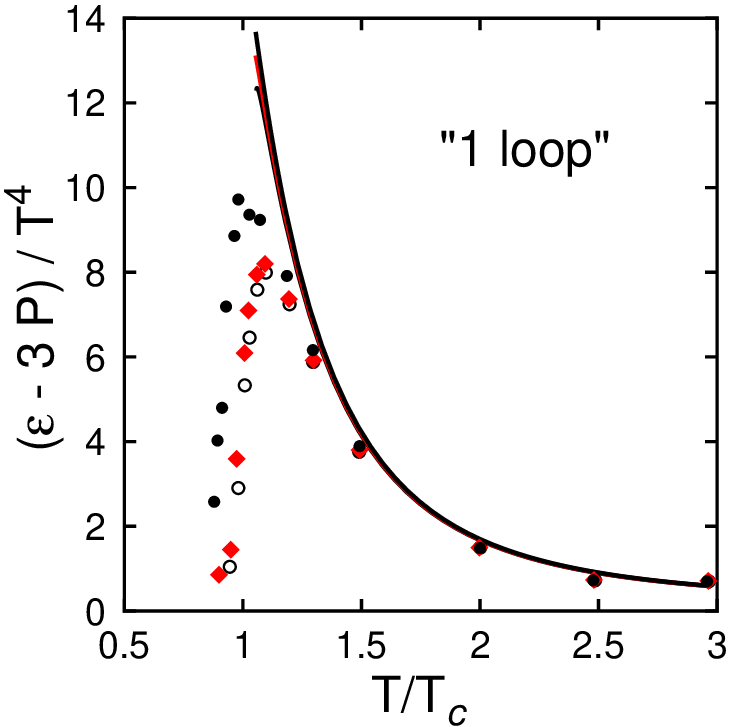}
\end{minipage}
\hspace*{9mm}
\begin{minipage}[h]{65mm} 
\includegraphics[height=64mm,clip]{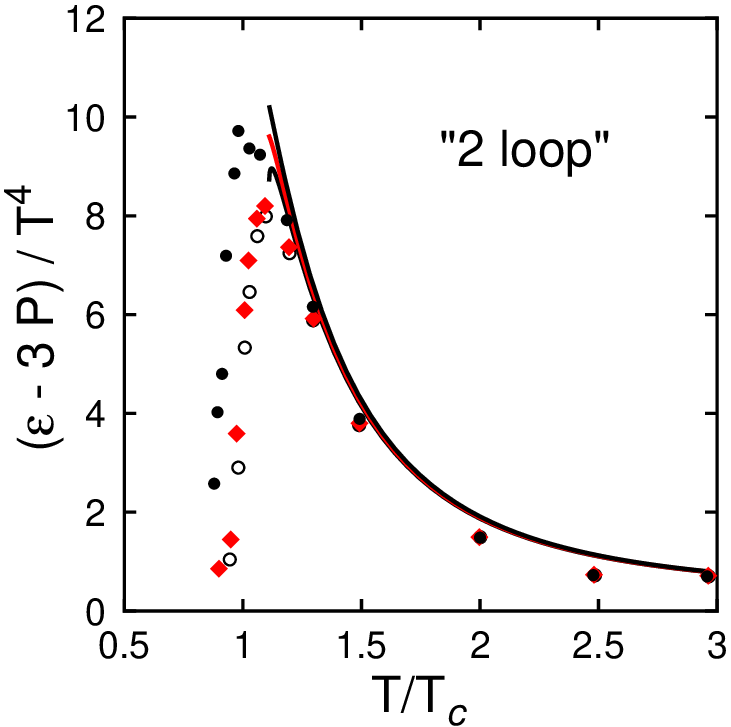}
\end{minipage}
\end{center}
\caption{Interaction measure, $\varepsilon -3P$, normalized by $T^4$
  as a function of $T/T_c$ at $\mu =$ 0, 330, 530 MeV (which are
  hardly distinguishable between each other) within the  ``1-loop''
  (left panel) and ``2-loop'' (right panel) calculations. The
  (2+1)-flavour lattice data 
  \cite{latEoS} for different  $\mu=$  0, 330, 530 MeV are displayed
  by open circles, diamonds and full circles, respectively, similarly
  to other figures. 
} \label{diff}
\end{figure}
\begin{figure}[h]
\begin{center}
\begin{minipage}[h]{65mm} 
\includegraphics[height=64mm,clip]{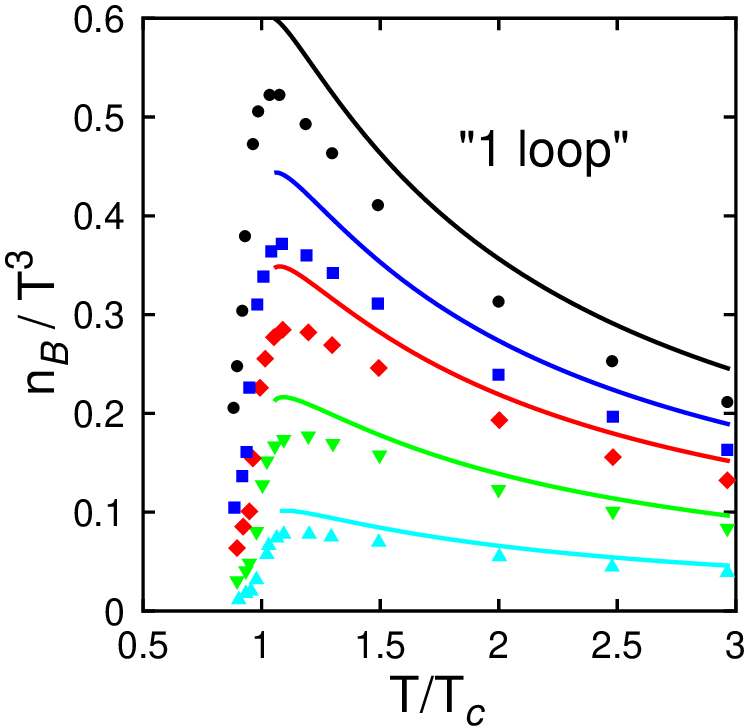}
\end{minipage}
\hspace*{9mm}
\begin{minipage}[h]{65mm} 
\includegraphics[height=64mm,clip]{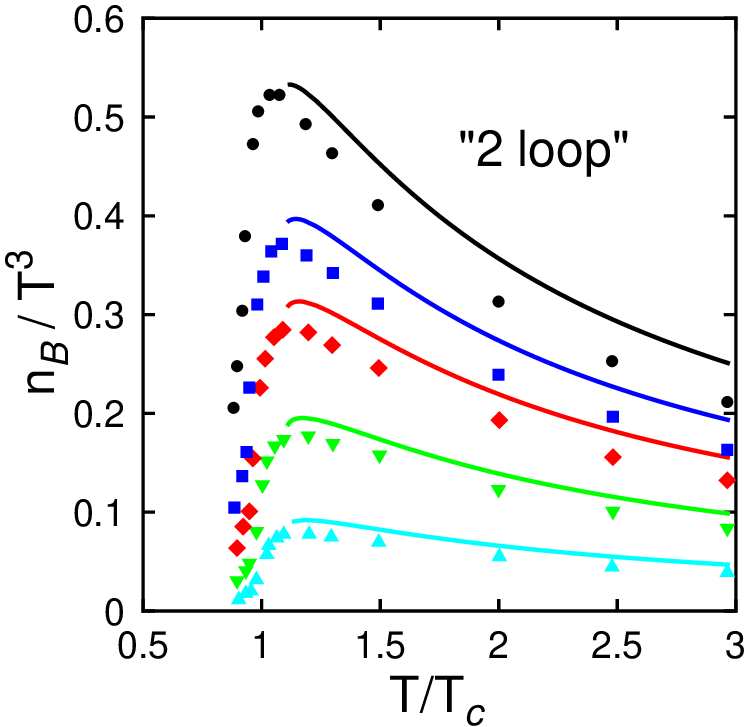}
\end{minipage}
\end{center}
\caption{Baryon density normalized by $T^3$ as a function of $T/T_c$
  at $\mu =$ 100, 210, 330, 410, 530 MeV (from bottom to top) within
  the  ``1-loop'' (left panel) and ``2-loop'' (right panel)
  calculations. The (2+1)-flavour lattice data are from \cite{latEoS}. 
} \label{nb}
\end{figure}

Now, when all the parameters are fixed to reproduce the lattice
pressure at $\mu=$ 0, all other calculations can be considered as
``predictions'' of the model. These results are presented in Figs
\ref{dp}--\ref{nb}. The model perfectly reproduces  
$\mu$-dependent part of the pressure, $\Delta P =
P(T,\mu)-P(T,\mu=0)$, Fig. \ref{dp}, and the ``interaction measure'',
$\varepsilon -3P$, Fig. \ref{diff}, at various chemical potentials. In
particular, it describes practical $\mu$-independence of the right
slope of $\varepsilon -3P$. At the same time, the lattice baryon
density, see Fig. \ref{nb}, turns out to be somewhat overestimated by
the model. In fact, this is not surprising, since the thermodynamic
consistency of continuum lattice limit is somewhat
unbalanced because of application of different normalization coefficients
to the raw lattice data: $c_p$  and $c_\mu$ \cite{latEoS} mentioned
above. Therefore, an exactly thermodynamically consistent model is 
unable to simultaneously reproduce all the continuum lattice data.

Note that the model fits the lattice quantities only above $T_C$. 
In view of arguments of Ref. \cite{Redlich} this is not
surprising. In \cite{Redlich} it is argued that below
$T_C$ these quantities are quantitatively well described by the resonance
hadronic gas. From this point of view, we cannot count on proper
description below $T_C$,  
since the hadronic degrees of freedom are completely disregarded by
the model.

On the other hand, we are able to reproduce the lattice data without
varying the overall normalization. However, for this we need
nontrivial auxiliary function $f(\chi)$. One of possible choices is  
\begin{eqnarray}
\label{2-loop}
f_{\scr{2-loop}}(\chi)=
1 + \arctan \left[\frac{\beta_1}{8\pi^2\beta_0}
g^2(\chi)\ln\frac{g^2(\chi)}{\lambda}\right]
\end{eqnarray}
with 
$$\lambda =  0.001\frac{16\pi^2}{\beta_0},$$
which we refer as ``2-loop'' choice, because with this
$f_{\scr{2-loop}}$ the coupling constant takes the 2-loop perturbative
limit at $T\to\infty$, cf. Eq. (\ref{g-0}). 
The additional $g^2 \ln (0.001)$ term is subleading as compared to
$g^2(\chi)\ln g^2(\chi)$ and thus does not prevent agreement with the
2-loop approximation for coupling constant. In fact, the function
$f_{\scr{2-loop}}$ is an ``exotic'' representation of a constant,
since in the temperature range under consideration, from $T_C$ to
$3T_C$, it is $f_{\scr{2-loop}}(\chi)\simeq$ 2.6 with good
accuracy. Precisely this enhancement of the coupling constant is
required to fit the actual overall normalization of the lattice
data. The reason of using function instead of the constant is only
that the function provides us with the  proper 2-loop asymptotic
limit. In this respect, any function $f$, providing us with additional
factor 2.6 in the temperature range from $T_C$ to $3T_C$ and
respecting the proper asymptotic limit of the coupling constant, is as
well suitable for this fit. 
The fitting procedure in this case is
completely similar to that for the ``1-loop'' choice. The obtained sets
of parameters for (2+1)- and 2-flavour cases are also summarized in
Table. The result of  fitting the pressure at $\mu=$ 0 is presented in
Fig. \ref{p0_fig}. The 2-flavour case requires here only slight tune
of the $B_C$ value as compared to the (2+1)-flavour case. 
``Predictions'' of the ``2-loop'' version are
demonstrated in Figs \ref{dp}--\ref{nb}. The quality of
reproduction of the lattice data here is approximately the same as in
the ``1-loop'' case.  

%
\begin{figure}[h]
\begin{center}
\begin{minipage}[h]{65mm} 
\includegraphics[height=64mm,clip]{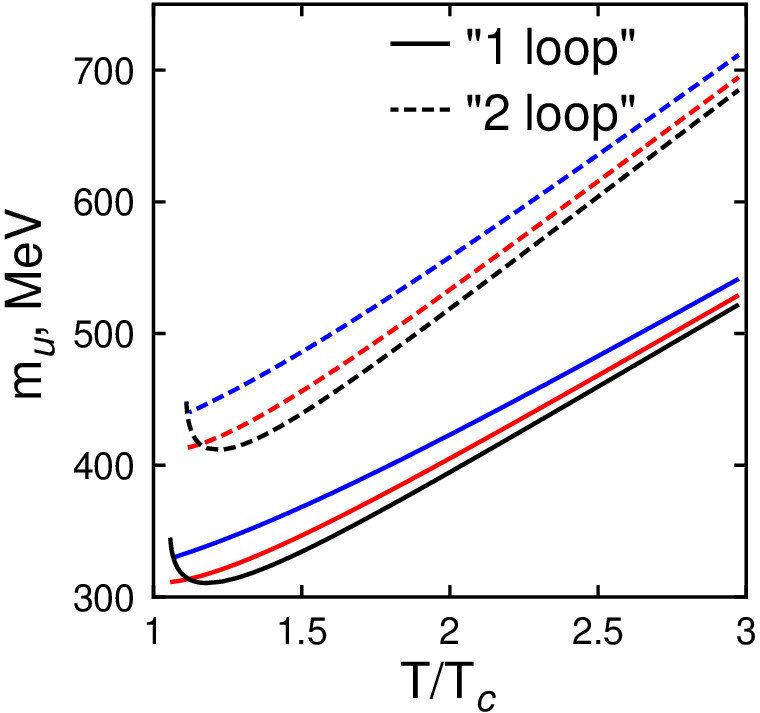}
\end{minipage}
\hspace*{9mm}
\begin{minipage}[h]{65mm} 
\includegraphics[height=64mm,clip]{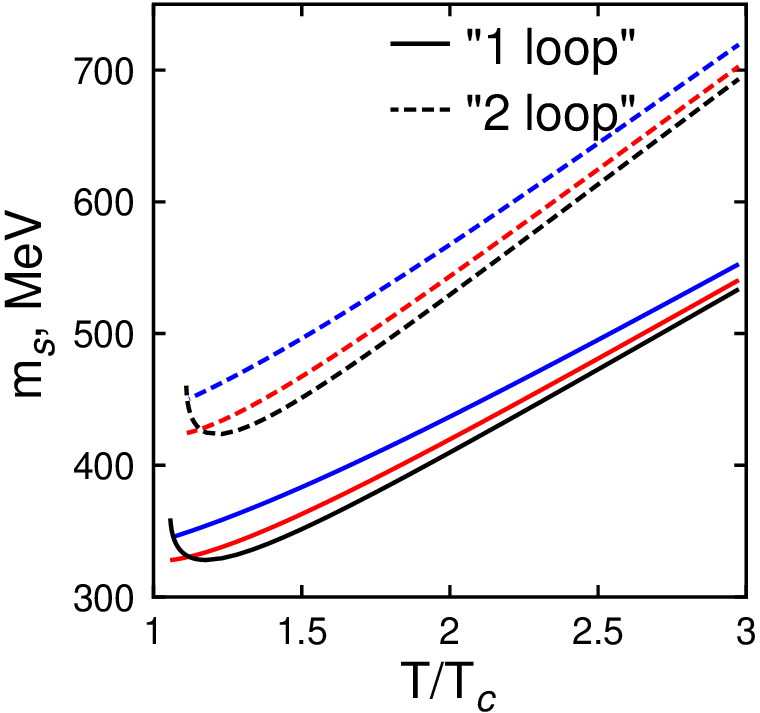}
\end{minipage}
\end{center}
\caption{Effective masses of $u$ and $d$ quarks (left panel) and $s$
  quark (right panel) as  functions of $T/T_c$ at $\mu =$ 0, 900,
  1500 MeV (from bottom to top). Solid and dashed lines correspond to
  ``1-loop'' and ``2-loop'' calculations, respectively.  
} \label{mass-q}
\end{figure}
\begin{figure}[t]
\begin{center}
\begin{minipage}[h]{65mm} 
\includegraphics[height=64mm,clip]{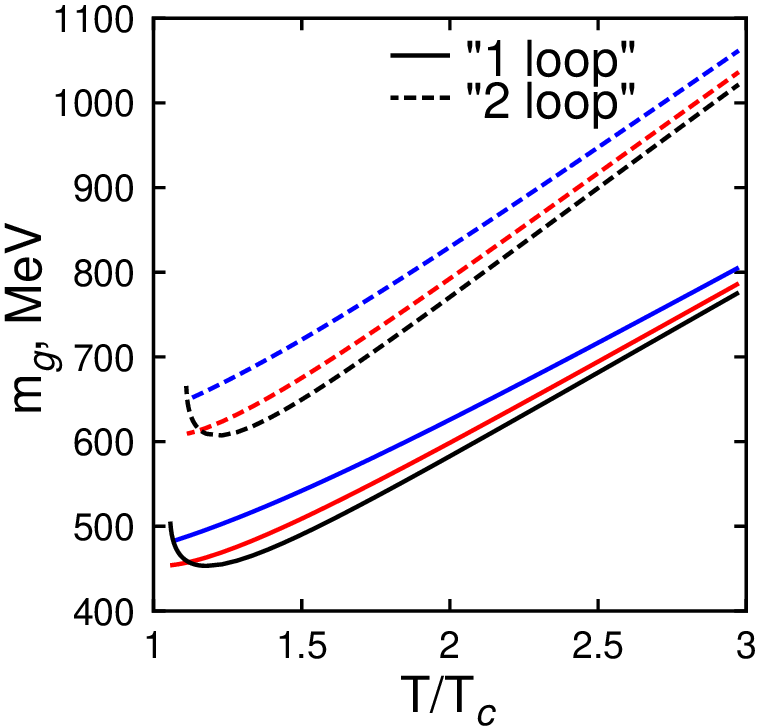}
\end{minipage}
\hspace*{9mm}
\begin{minipage}[h]{65mm} 
\includegraphics[height=64mm,clip]{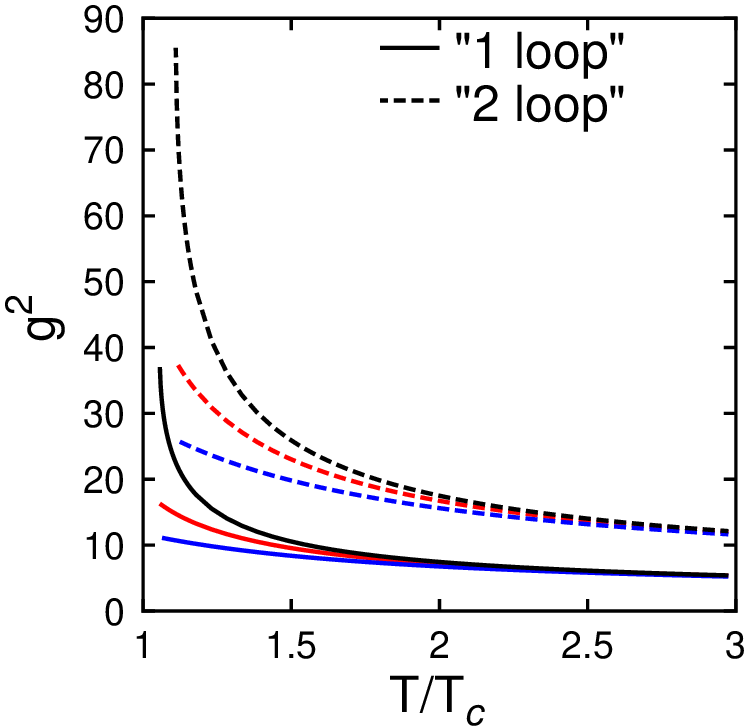}
\end{minipage}
\end{center}
\caption{Effective gluon mass (left panel) and coupling constant
  squared  (right panel) as a function of $T/T_c$ at $\mu =$ 0,
  900, 1500 MeV (from bottom to top in left panel, and from top to
  bottom in right panel). Solid and dashed lines
  correspond to ``1-loop'' and ``2-loop'' calculations, respectively.  
} \label{mass-g}
\end{figure}
%

In spite of the similar reproduction of lattice data, the two versions
of the model reveal quite different ``internal'' quantities, see Figs
\ref{mass-q} and \ref{mass-g}. Their absolute values differ by approximately
30\%, while the $T$ and $\mu$ dependences are very similar in the
``1-loop'' and ``2-loop'' versions. General trend of these dependences
is quite similar to those in the thermodynamic quasiparticle model
\cite{Szabo03}. Apparently, precisely this trend is essential for
reproduction of lattice data within both thermodynamic and present
quasiparticle models.

\section{Summary and Outlook}

We have presented a simple quasiparticle model aimed to interpret the
lattice QCD data. Similarly to existing quasiparticle approaches
\cite{Pesh96,Szabo03,Rebhan03,Weise01}, this model is motivated by the
lowest-order perturbative QCD. However, contrary to those models , it
is formulated in dynamical rather than thermodynamical terms. 
Presently we have succeeded only 
for the case $N_f\leq N_c$, where $N_f$ and $N_c$ are numbers of
quark flavours and colors, respectively. Nevertheless, this is quite a
general case for practical applications.  

The model has been applied to fit the lattice (2+1)-flavour QCD EoS at
finite baryon chemical potentials \cite{latEoS}. 
This is the most physical and important from the point
of view of practical applications case. However, we fragmentary
considered also 2-flavour case \footnote{Comparison with various
  lattice data on pure gauge and 2-flavour cases will be reported
  elsewhere.}. It is demonstrated
that a reasonable fit of the quark--gluon sector can be obtained with
different sets of 
phenomenological parameters. The ``1-loop'' version of
the model, cf. Eq. (\ref{1-loop}), certainly looks more natural, since it does
not involve an ``exotic'' auxiliary function $f$, as it does in the
``2-loop'' version, cf. Eq. (\ref{2-loop}). The only problem with the
``1-loop'' version is that it overestimates all lattice quantities by
approximately 10\% (and slightly more for the baryonic density). 
However, since the overall normalization of the lattice data is indeed
somewhat uncertain because of the poor extrapolation of these data to
the continuum limit, this misfit is quite acceptable.  

In spite of the difference in absolute values, ``internal'' quantities
of the model, like effective quark and gluon masses and coupling
constant, reveal very similar behaviour as functions of temperature
and chemical potential in both ``1-loop'' and ``2-loop'' versions.  
Moreover, this behaviour is also similar to that in thermodynamic
quasiparticle models  
\cite{Gorenstein95,Greiner,Levai,Pesh96,Szabo03,Rebhan03,Weise01}. 
Apparently,
precisely this general trend is essential for reproduction of lattice
data within both thermodynamic and present quasiparticle models.  

The presented model simulates the confinement of the QCD. In the
equilibrium case considered here, the solution to the model equations
simply does not exist below certain combination of the temperature and
the chemical potential. In particular, this is the reason why we are able
to fit the lattice quantities only above $T_C$. In \cite{Redlich} it
is argued that below $T_C$ these quantities are quantitatively
well described by the resonance hadronic gas. From this point of view,
we cannot count on proper description  below $T_C$, 
since the hadronic degrees of freedom are completely disregarded by
the model.   
From both theoretical and practical points of view, it is desirable to
include hadronic degrees of freedom in this model. Then we could count
on reproduction of EoS in the whole range of temperatures and chemical
potentials. Such a ``realistic'' EoS would be very useful in
hydrodynamic simulations of relativistic heavy-ion collisions.

\vspace*{5mm} {\bf Acknowledgements} \vspace*{5mm}

We are grateful to D.N. Voskresensky for fruitful discussions and
careful reading of the manuscript. 
This work was supported in part by the Deutsche
Forschungsgemeinschaft (DFG project 436 RUS 113/558/0-2), the
Russian Foundation for Basic Research (RFBR grant 03-02-04008) and 
Russian Minpromnauki (grant NS-1885.2003.2).


\begin{thebibliography}{99}
%
\bibitem{latEoS0}  F.~Karsch,
 Lect. Notes in Phys. {\bf 583}, 209 (2002),
[hep-lat/0106019];
AIP Conf. Proc. {\bf 602}, 323 (2001), [hep-lat/0109017]. 
%
\bibitem{latEoS} Z.~Fodor,
Nucl. Phys. {\bf A715}, 319 (2003), [hep-lat/0209101]; 
  F. Csikor, G.I. Egri, Z. Fodor, S.D. Katz, K.K.
Szabo, and A.I. Toth,
JHEP {\bf 405}, 46 (2004), [hep-lat/0401016].
%
\bibitem{Allton}
C.R. Allton, S. Ejiri, S.J. Hands, O. Kaczmarek, F. Karsch,
E. Laermann, and C. Schmidt,
Phys. Rev. D {\bf 68}, 014507 (2003), [hep-lat/0305007].
%
\bibitem{Gavai} R.V. Gavai and S. Gupta, Phys. Rev. D {\bf 68}, 034506
  (2003), [hep-lat/0303013]
%
\bibitem{Arnold} P. Arnold and C.X. Zhai, Phys. Rev. D {\bf 51}, 1906
  (1995); C.X. Zhai and B. Kastening, Phys. Rev. D {\bf 52}, 7232
  (1995). 
%
\bibitem{Pisarski} E. Braaten and R.D. Pisarski, Phys. Rev. D {\bf 45}, R1827
  (1992); J. Frenkel and J.C. Taylor, Nucl. Phys. {\bf B374}, 156
  (1992); J.P. Blaizot and E. Iancu, Nucl. Phys. {\bf B417}, 608
  (1994). 
%
\bibitem{Andersen} J.O. Andersen, E. Braaten, and M. Strickland,
  Phys. Rev. Lett. {\bf 83}, 2139 (1999), [hep-ph/9902327]; 
 Phys. Rev. D {\bf 62}, 045004 (2000), [hep-ph/0002048]; 
 J.O. Andersen, E. Braaten, E. Petitgirard, and M. Strickland,
  Phys. Rev. D {\bf 66}, 085016 (2002), [hep-ph/0205085]. 
%
\bibitem{Blaizot} 
J.P. Blaizot, E. Iancu, and A. Rebhan,
  Phys. Rev. Lett. {\bf 83}, 2906 (1999), [hep-ph/9906340]; 
Phys. Lett. {\bf B470}, 181 (1999), [hep-ph/9910309];
Phys. Rev. D {\bf 63}, 065003 (2001), [hep-ph/0005003].
%
\bibitem{Blaizot03} 
K. Kajantie, M. Laine, K. Rummukainen, and Y. Schr\"oder,
  Phys. Rev. D {\bf 67}, 105008 (2003), [hep-ph/0211321];
J.P. Blaizot, E. Iancu, and A. Rebhan,
Phys. Rev. D {\bf 68}, 025011 (2003), [hep-ph/0303045];  
A. Ipp, A. Rebhan and, A. Vuorinen, Phys. Rev D {\bf 69}, 077901 (2004),
[hep-ph/0311200]. 
%
\bibitem{Gorenstein95} M.I. Gorenstein and S.N. Yang, Phys. Rev. D
  {\bf 52}, 5206 (1995). 
%
\bibitem{Greiner} W. Greiner and D.H. Rischke, Phys. Rep. {\bf 264},
  183 (1996). 
%
\bibitem{Levai} P. Levai and U. Heinz, Phys. Rev. C {\bf 57}, 1879 
(1998), [hep-ph/9710463].
%
\bibitem{Pesh96}
A. Peshier, B. Kampfer, O.P. Pavlenko, and G. Soff, Phys. Rev. D {\bf 54},
2399 (1996); 
A. Peshier, B. Kampfer, and G. Soff, Phys. Rev. C {\bf 61}, 045203 (2000), 
[hep-ph/9911474];
A. Peshier, B. Kampfer, and G. Soff, Phys. Rev. D {\bf 66}, 094003 (2002)  
[hep-ph/0206229]; 
M. Bluhm, B. Kampfer, and G. Soff, hep-ph/0411106.
%
\bibitem{Szabo03} K.K. Szab\'o and A.I. T\'oth, JHEP {\bf 306}, 008
  (2003), [hep-ph/0302255]. 
%
\bibitem{Rebhan03} A. Rebhan and P. Romatschke, Phys. Rev. D {\bf 68},
  025022 (2003), [hep-ph/0304294]. 
%
\bibitem{Weise01} R.A. Schneider and W. Weise, Phys. Rev. C {\bf 64},
  055201 (2001), [hep-ph/0105242]; 
T. Renk, R.A. Schneider, and W. Weise, Phys. Rev. C {\bf 66}, 014902
(2002), [hep-ph/0201048]; 
M.A. Thaler, R. Schneider, and W. Weise, Phys.Rev. C {\bf 69}, 035210
(2004), [hep-ph/0310251]. 
%
\bibitem{Lebel96} M. Le Bellac, {\em Thermal Field Theory}, Cambridge
  University Press, Cambridge, 1996.
%
\bibitem{Kapusta}
J.I. Kapusta, {\em Finite-Temperature Field Theory}, Cambridge
University Press, Cambridge, 1989.
%
\bibitem{Yndurain}
F.J. Yndurain, {\em The Theory of Quark and Gluon Interactions},
Berlin,  Springer-Verlag, 1993. 
%
\bibitem{Redlich}  F.~Karsch, K. Redlich, and A. Tawfik,
  Eur. Phys. J. {\bf  C29}, 549 (2003), 
[hep-ph/0303108]. 
\end{thebibliography}
\end{document}